\documentclass[prl,twocolumn,showpacs,twoside,showpacs]{revtex4}

\usepackage{graphicx}
\usepackage{dcolumn}
\usepackage{bm}

\begin{document}

\title{Quantum fluctuations in the shape of exotic nuclei}
\author{J. M. Yao$^{1,2,4}$}
\author{J. Meng$^{2,3}$}
\author{P. Ring$^{4}$}
\author{Z. P. Li$^{2}$}
\author{K. Hagino$^{5}$}
 \affiliation{$^1$School of Physical Science and Technology,
                 Southwest University, Chongqing 400715, China}
 \affiliation{$^2$State Key Laboratory for Nuclear Physics and
                 Technology, School of Physics, Peking University, Beijing 100871, China}
 \affiliation{$^3$School of Physics and Nuclear Energy Engineering, Beihang University, Beijing 100191, China}
 \affiliation{$^4$Physik-Department der Technischen Universit\"at M\"unchen, D-85748 Garching, Germany}
 \affiliation{$^5$Department of Physics, Tohoku University, Sendai 980-8578, Japan}
 \date{\today}

\begin{abstract}
Quantum fluctuations concerning the shape of nuclei are treated
within the framework of covariant density functional theory. Long
range correlations beyond mean field are taken into account by
configuration mixing of wave functions with triaxial shapes and the
restoration of spontaneously broken rotational symmetries through
three-dimensional angular momentum projection. The {\it
controversial} nucleus $^{16}$C is treated as an example and it is
found that its ground state has a triaxial shape but with large shape
fluctuations. They are of crucial importance for a proper description
of the spectroscopic properties of such nuclei.
\end{abstract}

\pacs{21.10.Dr, 21.10.Re, 21.30.Fe, 21.60.Jz, 27.20.+n} \maketitle
\date{today}



Atomic nuclei are highly correlated and strongly interacting
many-body systems of mesoscopic character. On the one side they are
so small that quantum effects are dominant, on the other the number
of their constituents is finite and large enough so that classical
concepts such as their shape and orientation play an important role
for an understanding of their structure. Only in heavy nuclei are the
deformation parameters well defined quantities. In transitional
nuclei, and in light systems, quantum fluctuations have to be taken
into account.

The rapid development in radioactive nuclear beam facilities and
gamma ray detectors have in recent years allowed one to study exotic
nuclei far from stability and many new phenomena have been observed
and predicted in this context. In order to investigate the shape
degrees of freedom it is not sufficient to only study the ground
state properties of these nuclei, one also needs information about
the spectroscopy of the low-lying excited states. Therefore,
presently much interest is focused on the measurement of the energies
of the first $2^+$, or $4^+$ states and of the reduced transition
probabilities ($B(E2)$-values) from the first $2^+$ ($2^+_1$) to the
ground state ($0^+_1$)~\cite{MII.95x}. These are fundamental
quantities which reveal rich information about nuclear shapes and
shell structure.

Recently, the structure of the nucleus $^{16}$C having two neutron
particles and two proton holes above the doubly magic nucleus
$^{16}$O has become a very interesting and challenging topic. In
lifetime measurements using the recoil shadow method
(RSM)~\cite{Ong08} an anomalously hindered $B(E2 : 2_1^+ \rightarrow
0^+_1$) value of 2.6(9) e$^2$fm$^4$ was found, contradicting the low
excitation energy of E(2$^+_1$) = 1.766 MeV (see Ref.~\cite{Imai04}
for an earlier measurement). Furthermore, inelastic proton
scattering~\cite{Elekes04,Ong06} indicated a large quadrupole
deformation $\beta^{pp'}\sim$ 0.47(5). The quenched $B(E2)$ value,
combined with the large nuclear deformation, led to the suggestion
that the neutron motion plays a predominant role in the 2$^+_1$ state
of $^{16}$C~\cite{Elekes04,Ong06,Dombradi05}, although the structure
of $^{16}$C has not yet been fully understood. The more recent
lifetime measurement of the 2$^+_1$ state using the recoil distance
method (RDM) after a fusion-evaporation reaction gave a similar, but
slightly larger, $B(E2)$ value of 4.15(73)
e$^2$fm$^4$~\cite{Wiedeking08}.

On the theoretical side, very different models have been used to
describe the quenched $B(E2)$ value in $^{16}$C, including
antisymmetrized molecular dynamics (AMD)~\cite{Kanada97}, three-body
models~\cite{SMA.04x,HS.06x,Hagino07} and the shell
model~\cite{FMO.07x}. In AMD, the unusually small $B(E2)$ value
derived from the lifetime measurement with RSM was interpreted as the
coexistence of an oblate proton and a prolate neutron shape. In the
three-body models or in the shell model, a careful adjustment of an
effective charge or modifications of the Hamiltonian were required to
reproduce the data.

The self-consistent mean field approach derived from a global density
functional theory (DFT) provides a vivid way to study macroscopically
defined quantities, e.g. ground state energy, nuclear radius and
deformation. Exotic shapes of $^{16}$C have been investigated using
non-relativistic and relativistic DFT calculations, constrained by
deformation parameters~\cite{Zhang08,BGK.08x}. However, a very flat
energy surface has been found both on the prolate and oblate side
indicating that this is a transitional nucleus. Therefore,
substantial effects from con\-fi\-gu\-ra\-tion mixing connected with
shape fluctuations are expected which play an important role for
quantities of a quantum nature, e.g., discrete energy spectra and
transitions probabilities~\cite{Ring80}. In order to understand such
matrix elements, e.g. $B(E2)$ values, a microscopic approach going
beyond the mean-field level is required which is able to treat shape
quantum fluctuations properly.

In the framework of DFT, shape fluctuations and angular momentum
projection have been treated using non-relativistic density
functionals of Skyrme ~\cite{Valor00,Bender03} and
Gogny~\cite{Guzman02npa}, as well as covariant
functionals~\cite{Niksic06I}. In all of these cases, however,
intrinsic triaxiality was neglected which will be found to be crucial
for a reproduction of the data. Only very recently, a mixing of all
five quadrupole degrees of freedom has been attempted within the
context of triaxial Skyrme calculations~\cite{Bender08}.

The present investigation is based on a similar idea using covariant
density functional theory (CDFT). Based on Lorentz invariance, this
method connects the spin and spatial degrees of freedom of the
nucleus in a consistent way . Numerous investigations have shown that
the experimental data for the ground and excited states can be nicely
interpreted within a relativistic framework. Recently, we have
developed three-dimensional angular momentum projection (3DAMP) for
relativistic point coupling models to incorporate correlations
related to the restoration of broken rotational
symmetries~\cite{Yao09I}. This concept has now been further extended
to include fluctuations for triaxial deformations within the
framework of the Generator Coordinate Method (GCM).

In this letter, we use this method to study the effects of quantum
fluctuation for triaxial shapes in the {\it controversial} nucleus
$^{16}$C. It turns out that the ground state of $^{16}$C corresponds
to a triaxial shape having large shape fluctuations along the
$\gamma$ degree of freedom, which describes the triaxiality. Such a
new picture for the shape of $^{16}$C is essential to reproduce
quantitatively the experimental $B(E2: 2^+_1\rightarrow 0^+_1)$ value
of Ref.~\cite{Wiedeking08} that is consistent with the results from
inelastic ($p,p'$) scattering~\cite{Ong06}.


The set of deformed intrinsic wave functions $\vert q\rangle$ with
the quadrupole deformations in the Hill-Wheeler coordinates
$q=(\beta,\gamma)$~\cite{Ring80} is generated by constrained RMF+BCS
calculations using the parameter set PC-F1 of Ref.~\cite{BMM.02x}.
Further details are given in Ref.~\cite{Yao09I}. In particular, the
strength parameters of the zero-range pairing force are $V_n=308$ and
$V_p=321$ MeV$\cdot$fm$^3$.

\begin{figure}[t]
\vspace{-0.5cm}
 \includegraphics[width=8.2cm]{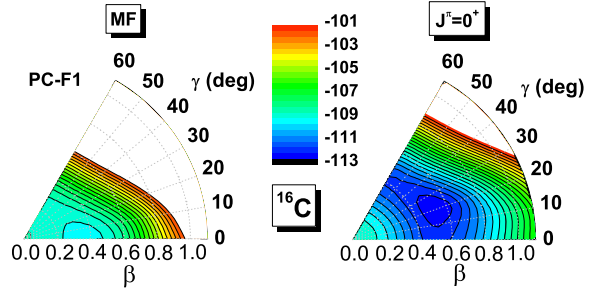}
 \caption{(Color online) Potential energy
 surfaces in $\beta$-$\gamma$ plane for the nucleus $^{16}$C
 obtained by triaxial relativistic mean-field calculations (left panel)
 and with projection onto angular momentum $J=0$ after the variation
 (right panel). The energy gap between two neighbor contour lines is
 $0.5$ MeV. }
 \label{PES}
 \end{figure}

In the left panel of Fig.~\ref{PES} we plot the potential energy
surface (PES) in $\beta$-$\gamma$ plane obtained with these mean
field calculations. This PES is rather soft, which is consistent with
the results found in Ref.~\cite{BGK.08x}. Of course, for all the
non-spherical points with ($\beta\neq0$) the intrinsic wave functions
$\vert\beta\neq0,\gamma\rangle$ have a certain orientation and
rotational symmetry is spontaneously broken. To restore this
symmetry, 3DAMP~\cite{Yao09I} is carried out and the right panel of
Fig.~\ref{PES} shows the projected PES for $J^\pi=0^+$.

\begin{figure*}[t]
 \centering
 \includegraphics[width=5.4cm]{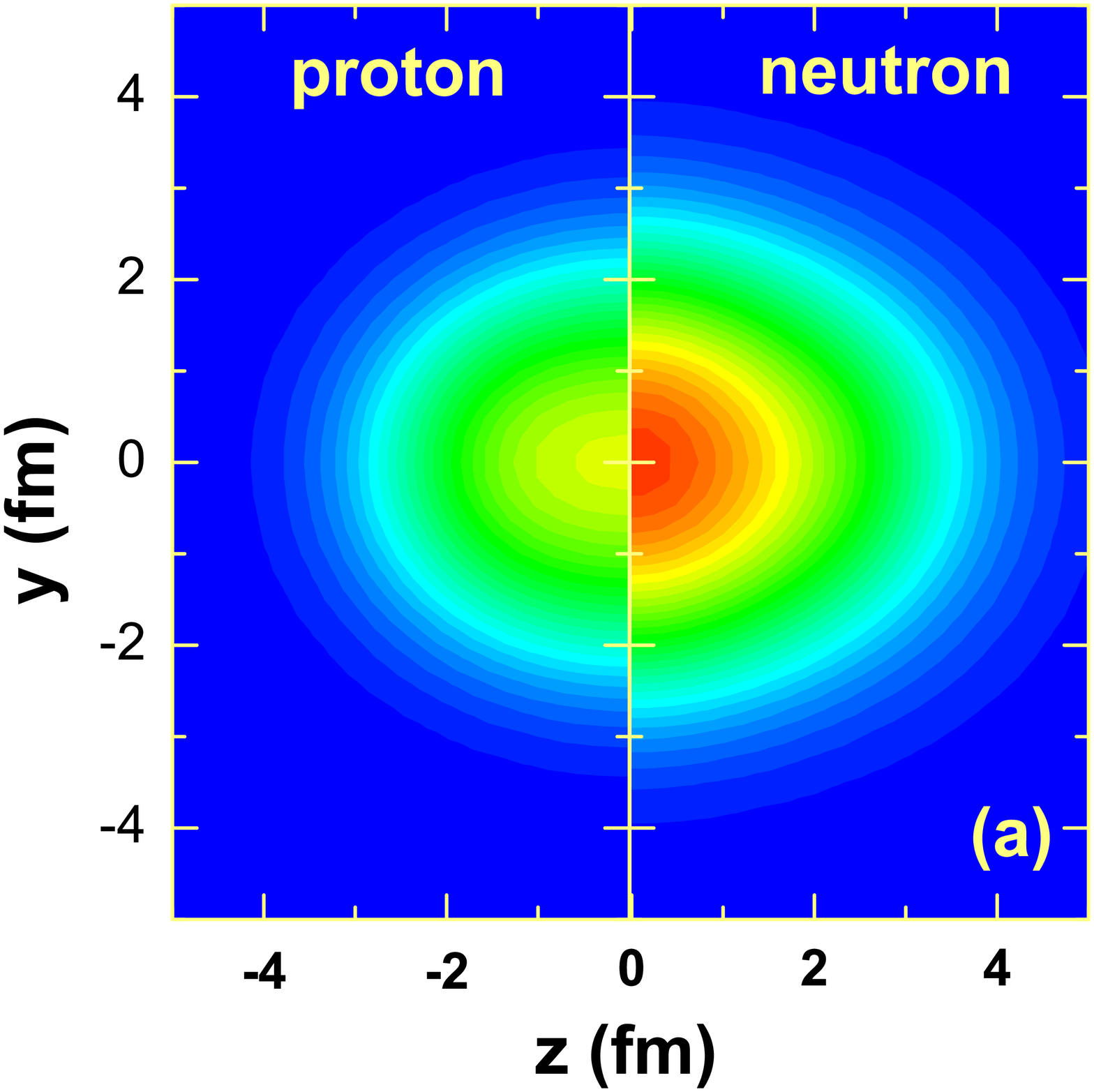}
 \includegraphics[width=5.4cm]{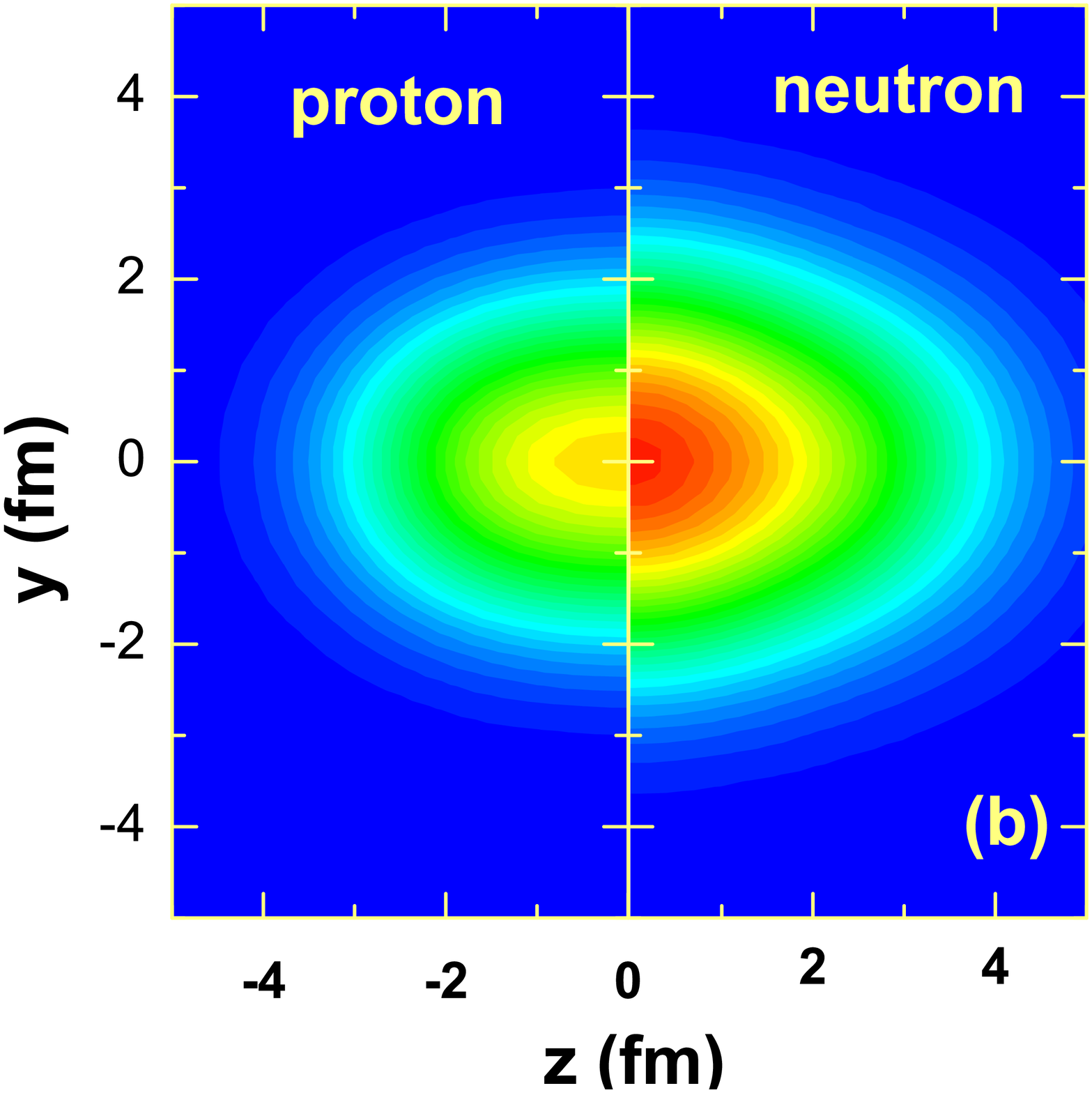}
 \includegraphics[width=5.4cm]{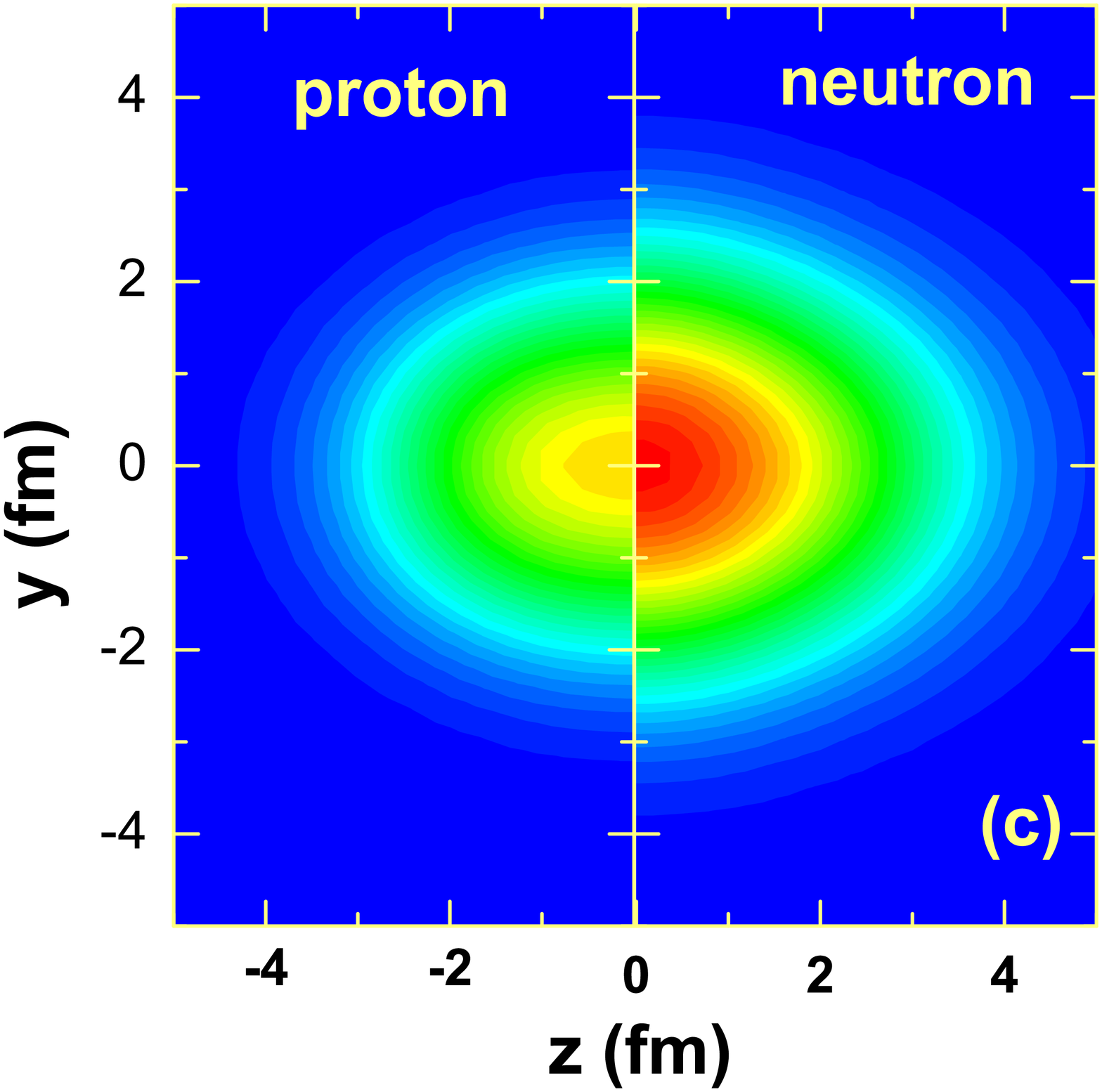}
\caption{(Color online) Intrinsic density distributions of protons
and neutrons in the $y$-$z$ plane (the $x$-axis has been integrated
over) (a) for the minimum of the mean-field potential energy surface
(PES) of Fig.~\ref{PES}a at $\beta=0.4, \gamma=0^\circ$; (b) for the
minimum of the projected PES ($0^+$) in Fig.~\ref{PES}b with
$\beta=0.6, \gamma=20^\circ$; (c) for the deformation of the
projected GCM state $0^+_1$ with $\langle\beta\rangle=0.44,
\langle\gamma\rangle=24.27^\circ$.} \label{Density}
\end{figure*}

Of particular interest is the occurrence of an obvious triaxial
minimum with $\beta=0.6, \gamma\sim20^\circ$. To demonstrate the
shapes at the various points in the ($\beta$-$\gamma$)-plane, in
Fig.~\ref{Density} we plot the corresponding intrinsic density
distributions for neutrons and protons in the $y$-$z$ plane
integrated over the $x$-coordinate. Fig.~\ref{Density}a refers to the
axially symmetric minimum in the unprojected PES of Fig.~\ref{PES}a
and Fig.~\ref{Density}b corresponds to the minimum in the projected
PES of Fig.~\ref{PES}b. Here we observe rather different deformation
parameters for neutrons ($\beta_n=0.69, \gamma_n=15.48^\circ$) and
protons ($\beta_p=0.46, \gamma_p=31.47^\circ$), which shows us
another kind of decoupled structure for the density distribution of
neutron and proton, different to that indicated in
Ref.~\cite{Kanada97}. A careful analysis shows that the difference
between neutron and proton deformation is due to the special 2p-2h
configurations $\nu (1d_{5/2})^2\otimes\pi (1p_{3/2})^{-2}$. For
large prolate deformations ($\beta>1.2$), the deformation driving
orbit $1d_{5/2}$ dives into the Fermi sea of protons as well. In this
case, neutrons and protons have almost the same deformation and the
decoupled structure ceases.

Although there is an obvious minimum in the projected PES shown in
Fig.~\ref{PES}, this minimum is still very soft along the $\gamma$
direction. One has, therefore, to allow for superpositions of wave
functions with different deformations and to perform a GCM
calculation in the full $\beta$-$\gamma$ plane
(for details see Ref.~\cite{Yao09I})%
\begin{equation}
\vert\Psi^{J}_M\rangle=\sum_K\int d^2q f_K(q) P^{J}_{MK}\vert
q\rangle
\end{equation}%
including correlations due to restoration of broken symmetries and
fluctuations of the deformation coordinates.

\begin{figure}[h!]
 \vspace{-0.8cm}
 \centering
 \includegraphics[width=7cm]{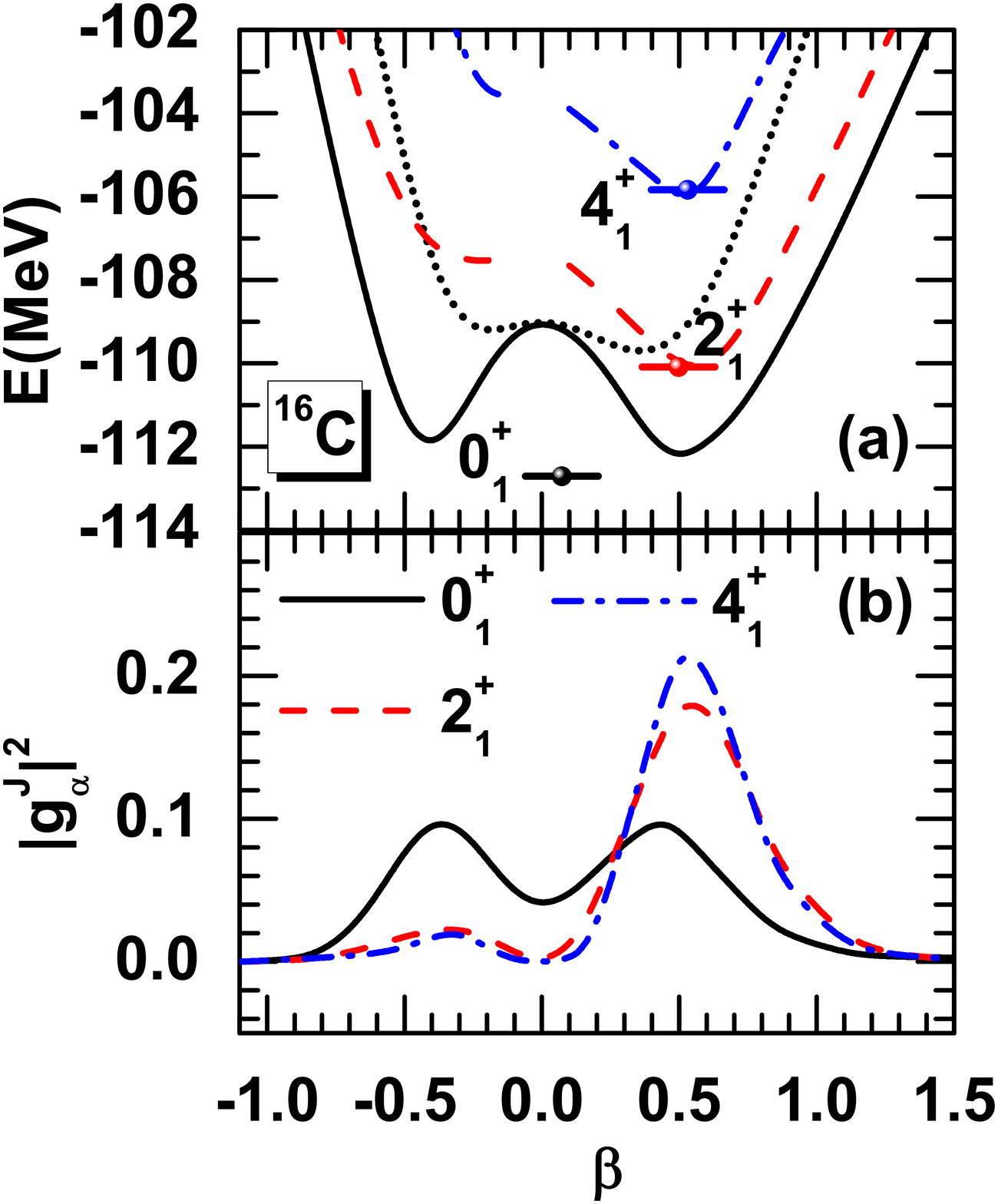} \vspace{-0.5cm}
 \caption{(Color online) Results of 1DAMP+GCM calculation with axially symmetric intrinsic
  states:  (a) energies and average quadrupole moments of
  the lowest GCM states for  $J^\pi=0^+, 2^+, 4^+$ in $^{16}$C,
  together with the mean-field (dotted line) and projected energy curves.
  (b) squares of collective wave functions $\vert g^J_\alpha(\beta)\vert^2$
  for the corresponding lowest GCM states.
}
 \label{1DAMPGCM}
\end{figure}

To illustrate the importance of shape fluctuations in the $\gamma$
degree of freedom, in a first step, we restrict ourselves to axially
symmetric shapes and choose the generator coordinates as $\beta$=$0,
0.1, \cdots, 1.5$ for $\gamma$=$0^\circ,$ and $\gamma$=$180^\circ$.
As the result, the full 3DAMP+GCM calculation is simplified into a
1DAMP+GCM calculation and all components with $K\neq0$ vanish. In
Fig. \ref{1DAMPGCM}a we plot energies and average quadrupole moments
of the lowest GCM states for each angular momentum $0^+, 2^+, 4^+$
together with the mean-field PES and the projected PES as a function
of $\beta$.  The corresponding probability distributions $\vert
g^J_\alpha(\beta)\vert^2$ are shown in Fig. \ref{1DAMPGCM}b where
$g^{J}_\alpha(\beta)$ is the solution of the Hill-Wheeler-Griffin
equation in the "natural basis" (for details see Ref.~\cite{Ring80}).

Fig.~\ref{1DAMPGCM} shows two minima on the projected PES with
$J^\pi=0^+$, one on the prolate and one on the oblate side having
similar probabilities. Therefore, 1DAMP+GCM calculations lead to a
nearly vanishing quadrupole moment for the ground state with a rather
small $B(E2:2^+_1\rightarrow 0^+_1)=0.11$e$^2$fm$^4$. According to
Fig.~\ref{PES}, it is essential to include triaxial states and to
carry out a full 3DAMP+GCM calculation. The probabilities in
Fig.~\ref{1DAMPGCM}b decay quickly for $\beta>1.1$. It is therefore
justified to choose triaxial states with $0\leq\beta\leq1.1$ and
$0\leq\gamma\leq60^\circ$ having steps $\Delta\beta=0.1$ and
$\Delta\gamma=10^\circ$ respectively.

\begin{figure}[t]
 \centering
 \includegraphics[width=9.5cm]{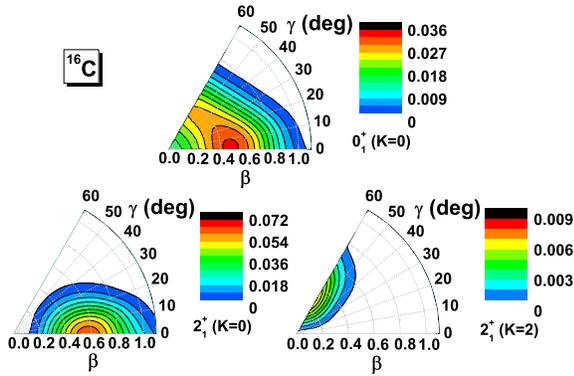}
 \caption{(Color online) Contour plots of the probability distributions
 in $\beta$ and $\gamma$ from 3DAMP+GCM
 calculation for the ground state 
 and the first excited state 
 in $^{16}$C.}
 \label{3DAMPGCM}
\end{figure}

Fig. \ref{3DAMPGCM} shows contour plots of the probability
distributions in the $\beta$-$\gamma$ plane resulting from 3DAMP+GCM
calculations for both the ground state ($0^+_1$) and the first
excited state ($2^+_1$). The ground state has a triaxial structure
with almost uniform probability along the $\gamma$-direction which
indicates an obvious quantum shape fluctuation in $\gamma$ direction.
For $J=2$ we have $K$-mixing and we therefore show the two
distributions for $K=0$ and $K=2$ separately. Both are concentrated
along the axially symmetric configurations with $K=0$ on the prolate
and $K=2$ on the oblate side. However, the $K=0$ part exhausts 93.1\%
of the norm so that the nucleus has a strong prolate deformation in
the ($2^+_1$) state with $\beta\simeq 0.6$.

\begin{table}[]
\vspace{-0.5cm}%
\centering%
\tabcolsep=3pt%
\renewcommand{\arraystretch}{1.5}%
\caption{Average deformation parameters $\langle \beta\rangle$,
$\langle \gamma\rangle$ for the ground state ($0^+_1$) obtained from
3DAMP+GCM calculations. The quantities $\beta^{J=0}_{\rm min},
\gamma^{J=0}_{\rm min}$ and $\beta^{\rm MF}_{\rm min}, \gamma^{\rm
MF}_{\rm min}$ are the deformations at the minima of the projected
($J=0$) and mean-field potential energy surfaces.}%
\vspace{0.3cm}
\begin{tabular}{ccccccc}
\hline\hline%
&~~~$\langle\beta\rangle$~~&~~$\langle \gamma\rangle$~~~&%
$~~~~\beta^{J=0}_{\rm min}$~&~$\gamma^{J=0}_{\rm min}$~~~&%
$~~\beta^{\rm MF}_{\rm min}$~&~$\gamma^{\rm MF}_{\rm min}$~~\\%
\hline%
neut. &~~~0.50~&~21.41$^\circ$~&~~~0.69~&~15.48$^\circ$~&~~0.48~~&~~0$^\circ$ \\%
prot. &~~~0.34~&~31.30$^\circ$~&~~~0.46~&~31.47$^\circ$~&~~0.27~~&~~0$^\circ$ \\%
total &~~~0.44~&~24.27$^\circ$~&~~~0.60~&~20.00$^\circ$~&~~0.40~~&~~0$^\circ$ \\%
\hline\hline%
\end{tabular}
\label{tab1}
\end{table}

For comparison with the mean-field results, we define the quadrupole
moments for the ground state
\begin{equation}
\langle Q_{2\mu}\rangle \equiv \int d^2q |g^{J=0}_{\alpha=1}(q)|^2
q_{2\mu}(q),\qquad \mu=0,2.
\end{equation}
The deformations $\displaystyle\langle \beta\rangle=4\pi\sqrt{\langle
Q_{20}\rangle^2+2\langle Q_{22}\rangle^2}/(3NR^2_0)$ and
$\displaystyle\tan\langle \gamma\rangle=\sqrt{2}\langle
Q_{22}\rangle/\langle Q_{20}\rangle$ shown in Tab.~\ref{tab1} result
from 3DAMP+GCM calculations. We observe a decoupled structure, where
the average neutron deformation $\langle\beta\rangle_n=0.50$ is
obviously larger than that of protons $\langle\beta\rangle_p=0.34$.
This can also be seen in Fig.~\ref{Density} where
\begin{equation}
 \label{Density_GCM}
  \rho^{GCM}_\tau(\bm{r})
  \equiv \int d^2q |g^{J=0}_{\alpha=1}(q)|^2\rho_\tau(\bm{r};q)
\end{equation}
are the density distributions of neutrons and protons for the
projected GCM state ($0^+_1$).

\begin{figure}[t]
 \centering
 \includegraphics[width=8.5cm]{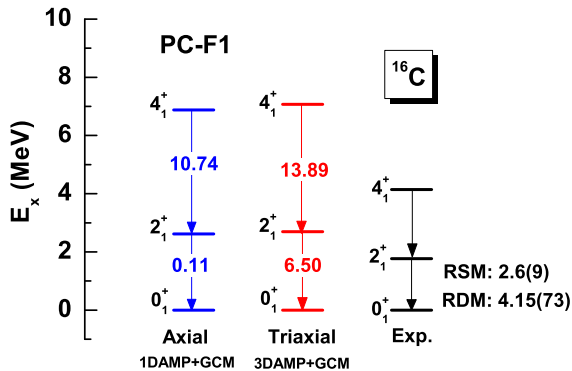} 
 \caption{(Color online) The lowest energy levels of angular momentum $J=0,2,4$ in $^{16}$C. The
 $B(E2)$ values are in units of e$^2$fm$^4$. The experimental data are taken from
 Refs.~\cite{Ong08,Wiedeking08}. }
 \label{Spectrum}
 \end{figure}

To demonstrate the importance of shape fluctuations in $\gamma$ for
spectroscopic properties, we plot in Fig.~\ref{Spectrum} the lowest
energy levels having angular momentum $J=0,2,4$ in $^{16}$C obtained
from the 1D and 3D AMP+GCM calculations and compare them with
experiment~\cite{Wiedeking08}. The inclusion of triaxial states does
not change the energy spectrum too much, but it improves the electric
quadrupole transitions significantly. Only with a proper treatment of
the shape fluctuations in $\gamma$, are the calculated $B(E2:
2^+_1\rightarrow 0^+_1)=6.50$ e$^2$fm$^4$ in agreement with recent
data.

The predicted ratio $R_{4/2}\equiv E_4/E_2=2.63$ is a little larger
than the experimental value $R_{4/2}=2.35$. Both are close to the
$R_{4/2}=2.50$, a typical value for $\gamma$-soft nuclei. The
predicted deviations of $R_{4/2}$ from experiment go in the direction
of a rotor ($R_{4/2}=3$). This has its origin in the increasing
prolate deformation as angular momentum increases to $J=2$ and $J=4$.
As observed in many calculations using projection after
variation~\cite{Valor00,Bender03,Guzman02npa,Niksic06I} the
calculated spectrum is systematically stretched. It would be far more
compressed by the inclusion of a cranking term in the mean-field
calculation~\cite{ZSDK.07x}.

In summary, starting from covariant density functional theory and
including additional correlations by restoring symmetries and by
configuration mixing of triaxial shapes we have studied the influence
of quantum shape fluctuations on spectroscopic properties in the
exotic nucleus $^{16}$C. In contrary to earlier investigations using
other methods we find that the ground state has a triaxial shape with
large shape fluctuations in $\gamma$. Such a novel picture for the
shape of $^{16}$C is essential to reproduce quantitatively
$B(E2\downarrow)$ values derived from a recent lifetime measurement
based on the RDM method.

This work has been supported by the Major State Basic Research
Developing Program 2007 CB815000, the Asia-Europe Link Project
[CN/ASIA-LINK/008 (094-791)] of the EU,
the NSFC under Grant Nos. 10775004, 10705004, the DFG Cluster of
Excellence ``Origin and Structure of the Universe",
and by MEXT KAKENHI 19740115. We thank R. R. Hilton and D. Vretenar
for helpful discussions.


\begin{thebibliography}{23}
\expandafter\ifx\csname
natexlab\endcsname\relax\def\natexlab#1{#1}\fi
\expandafter\ifx\csname bibnamefont\endcsname\relax
  \def\bibnamefont#1{#1}\fi
\expandafter\ifx\csname bibfnamefont\endcsname\relax
  \def\bibfnamefont#1{#1}\fi
\expandafter\ifx\csname citenamefont\endcsname\relax
  \def\citenamefont#1{#1}\fi
\expandafter\ifx\csname url\endcsname\relax
  \def\url#1{\texttt{#1}}\fi
\expandafter\ifx\csname urlprefix\endcsname\relax\def\urlprefix{URL
}\fi \providecommand{\bibinfo}[2]{#2}
\providecommand{\eprint}[2][]{\url{#2}}

\bibitem[{\citenamefont{Motobayashi~{\it et al.}}(1995)}]{MII.95x}
\bibinfo{author}{\bibfnamefont{T.}~\bibnamefont{Motobayashi~{\it et al.}}},
  \bibinfo{journal}{Phys. Lett.} \textbf{\bibinfo{volume}{B346}},
  \bibinfo{pages}{9} (\bibinfo{year}{1995}).

\bibitem[{\citenamefont{Ong~{\it el al}.}(2008)}]{Ong08}
\bibinfo{author}{\bibfnamefont{H.~J.} \bibnamefont{Ong~{\it el al}.}},
  \bibinfo{journal}{Phys. Rev.} \textbf{\bibinfo{volume}{C78}},
  \bibinfo{pages}{014308} (\bibinfo{year}{2008}).

\bibitem[{\citenamefont{Imai~{\it et al}.}(2004)}]{Imai04}
\bibinfo{author}{\bibfnamefont{N.}~\bibnamefont{Imai~{\it et al}.}},
  \bibinfo{journal}{Phys. Rev. Lett.} \textbf{\bibinfo{volume}{92}},
  \bibinfo{pages}{062501} (\bibinfo{year}{2004}).

\bibitem[{\citenamefont{Elekes~{\it et al}.}(2004)}]{Elekes04}
\bibinfo{author}{\bibfnamefont{Z.}~\bibnamefont{Elekes~{\it et al}.}},
  \bibinfo{journal}{Phys. Lett.} \textbf{\bibinfo{volume}{B586}},
  \bibinfo{pages}{34} (\bibinfo{year}{2004}).

\bibitem[{\citenamefont{Ong~{\it el al}.}(2006)}]{Ong06}
\bibinfo{author}{\bibfnamefont{H.~J.} \bibnamefont{Ong~{\it el al}.}},
  \bibinfo{journal}{Phys. Rev.} \textbf{\bibinfo{volume}{C73}},
  \bibinfo{pages}{024610} (\bibinfo{year}{2006}).

\bibitem[{\citenamefont{Zs. Dombradi~{\it et al}}(2005)}]{Dombradi05}
\bibinfo{author}{\bibfnamefont{Z.}~\bibnamefont{Zs. Dombradi~{\it et al}}},
  \bibinfo{journal}{Phys. Lett.} \textbf{\bibinfo{volume}{B621}},
  \bibinfo{pages}{81} (\bibinfo{year}{2005}).

\bibitem[{\citenamefont{Wiedeking~{\it el al}.}(2008)}]{Wiedeking08}
\bibinfo{author}{\bibfnamefont{M.}~\bibnamefont{Wiedeking~{\it el al}.}},
  \bibinfo{journal}{Phys. Rev. Lett.} \textbf{\bibinfo{volume}{100}},
  \bibinfo{pages}{152501} (\bibinfo{year}{2008}).

\bibitem[{\citenamefont{Kanada-En'yo and Horiuchi}(1997)}]{Kanada97}
\bibinfo{author}{\bibfnamefont{Y.}~\bibnamefont{Kanada-En'yo}}
  \bibnamefont{and} \bibinfo{author}{\bibfnamefont{H.}~\bibnamefont{Horiuchi}},
  \bibinfo{journal}{Phys. Rev.} \textbf{\bibinfo{volume}{C55}},
  \bibinfo{pages}{2860} (\bibinfo{year}{1997}).

\bibitem[{\citenamefont{Suzuki~{\it et al.}}(2004)}]{SMA.04x}
\bibinfo{author}{\bibfnamefont{Y.}~\bibnamefont{Suzuki~{\it et al.}}},
  \bibinfo{journal}{Phys. Rev.} \textbf{\bibinfo{volume}{C70}},
  \bibinfo{pages}{051302(R)} (\bibinfo{year}{2004}).

\bibitem[{\citenamefont{Horiuchi~{\it et al,}}(2006)}]{HS.06x}
\bibinfo{author}{\bibfnamefont{W.}~\bibnamefont{Horiuchi~{\it et al,}}},
  \bibinfo{journal}{Phys. Rev.} \textbf{\bibinfo{volume}{C73}},
  \bibinfo{pages}{037304} (\bibinfo{year}{2006}).

\bibitem[{\citenamefont{Hagino and Sagawa}(2007)}]{Hagino07}
\bibinfo{author}{\bibfnamefont{K.}~\bibnamefont{Hagino}} \bibnamefont{and}
  \bibinfo{author}{\bibfnamefont{H.}~\bibnamefont{Sagawa}},
  \bibinfo{journal}{Phys. Rev.} \textbf{\bibinfo{volume}{C75}},
  \bibinfo{pages}{021301(R)} (\bibinfo{year}{2007}).

\bibitem[{\citenamefont{Fujii~{\it et al.}}(2007)}]{FMO.07x}
\bibinfo{author}{\bibfnamefont{S.}~\bibnamefont{Fujii~{\it et al.}}},
  \bibinfo{journal}{Phys. Lett.} \textbf{\bibinfo{volume}{B650}},
  \bibinfo{pages}{9} (\bibinfo{year}{2007}).

\bibitem[{\citenamefont{Zhang et~al.}(2008)\citenamefont{Zhang, Sagawa,
  Yoshino, Hagino, and Meng}}]{Zhang08}
\bibinfo{author}{\bibfnamefont{Y.}~\bibnamefont{Zhang}},
  \bibinfo{author}{\bibfnamefont{H.}~\bibnamefont{Sagawa}},
  \bibinfo{author}{\bibfnamefont{D.}~\bibnamefont{Yoshino}},
  \bibinfo{author}{\bibfnamefont{K.}~\bibnamefont{Hagino}}, \bibnamefont{and}
  \bibinfo{author}{\bibfnamefont{J.}~\bibnamefont{Meng}},
  \bibinfo{journal}{Prog. Theor. Phys.} \textbf{\bibinfo{volume}{120}},
  \bibinfo{pages}{129} (\bibinfo{year}{2008}).

\bibitem[{\citenamefont{B\"{u}rvenich~{\it et al.}}(2008)}]{BGK.08x}
\bibinfo{author}{\bibfnamefont{T.}~\bibnamefont{B\"{u}rvenich~{\it et al.}}},
  \bibinfo{journal}{J. Phys.} \textbf{\bibinfo{volume}{G35}},
  \bibinfo{pages}{025103} (\bibinfo{year}{2008}).

\bibitem[{\citenamefont{Ring and Schuck}(1980)}]{Ring80}
\bibinfo{author}{\bibfnamefont{P.}~\bibnamefont{Ring}} \bibnamefont{and}
  \bibinfo{author}{\bibfnamefont{P.}~\bibnamefont{Schuck}},
  \emph{\bibinfo{title}{The Nuclear Many-Body Problem}}
  (\bibinfo{publisher}{Springer}, \bibinfo{address}{Heidelberg},
  \bibinfo{year}{1980}).

\bibitem[{\citenamefont{Valor et~al.}(2000)\citenamefont{Valor, Heenen, and
  Bonche}}]{Valor00}
\bibinfo{author}{\bibfnamefont{A.}~\bibnamefont{Valor}},
  \bibinfo{author}{\bibfnamefont{P.-H.} \bibnamefont{Heenen}},
  \bibnamefont{and} \bibinfo{author}{\bibfnamefont{P.}~\bibnamefont{Bonche}},
  \bibinfo{journal}{Nucl. Phys.} \textbf{\bibinfo{volume}{A671}},
  \bibinfo{pages}{145} (\bibinfo{year}{2000}).

\bibitem[{\citenamefont{Bender et~al.}(2003)\citenamefont{Bender, Heenen, and
  Reinhard}}]{Bender03}
\bibinfo{author}{\bibfnamefont{M.}~\bibnamefont{Bender}},
  \bibinfo{author}{\bibfnamefont{P.-H.} \bibnamefont{Heenen}},
  \bibnamefont{and} \bibinfo{author}{\bibfnamefont{P.-G.}
  \bibnamefont{Reinhard}}, \bibinfo{journal}{Rev. Mod. Phys.}
  \textbf{\bibinfo{volume}{75}}, \bibinfo{pages}{121} (\bibinfo{year}{2003}).

\bibitem[{\citenamefont{Rodr{\'i}guez-Guzm\'{a}n
  et~al.}(2002)\citenamefont{Rodr{\'i}guez-Guzm\'{a}n, Egido, and
  Robledo}}]{Guzman02npa}
\bibinfo{author}{\bibfnamefont{R.}~\bibnamefont{Rodr{\'i}guez-Guzm\'{a}n}},
  \bibinfo{author}{\bibfnamefont{J.~L.} \bibnamefont{Egido}}, \bibnamefont{and}
  \bibinfo{author}{\bibfnamefont{L.~M.} \bibnamefont{Robledo}},
  \bibinfo{journal}{Nucl. Phys.} \textbf{\bibinfo{volume}{A709}},
  \bibinfo{pages}{201} (\bibinfo{year}{2002}).

\bibitem[{\citenamefont{Nik\v{s}i\'{c}
  et~al.}(2006)\citenamefont{Nik\v{s}i\'{c}, Vretenar, and Ring}}]{Niksic06I}
\bibinfo{author}{\bibfnamefont{T.}~\bibnamefont{Nik\v{s}i\'{c}}},
  \bibinfo{author}{\bibfnamefont{D.}~\bibnamefont{Vretenar}}, \bibnamefont{and}
  \bibinfo{author}{\bibfnamefont{P.}~\bibnamefont{Ring}},
  \bibinfo{journal}{Phys. Rev.} \textbf{\bibinfo{volume}{C73}},
  \bibinfo{pages}{034308} (\bibinfo{year}{2006}).

\bibitem[{\citenamefont{Bender and Heenen}(2008)}]{Bender08}
\bibinfo{author}{\bibfnamefont{M.}~\bibnamefont{Bender}} \bibnamefont{and}
  \bibinfo{author}{\bibfnamefont{P.-H.} \bibnamefont{Heenen}},
  \bibinfo{journal}{Phys. Rev.} \textbf{\bibinfo{volume}{C78}},
  \bibinfo{pages}{024309} (\bibinfo{year}{2008}).

\bibitem[{\citenamefont{Yao et~al.}(2009)\citenamefont{Yao, Meng, Ring, and
Pena Arteaga}}]{Yao09I}
\bibinfo{author}{\bibfnamefont{J.~M.} \bibnamefont{Yao}},
  \bibinfo{author}{\bibfnamefont{J.}~\bibnamefont{Meng}},
  \bibinfo{author}{\bibfnamefont{P.}~\bibnamefont{Ring}}, \bibnamefont{and}
  \bibinfo{author}{\bibfnamefont{D.~} \bibnamefont{Pe\~na Arteaga}},
  \bibinfo{journal}{Phys. Rev.} \textbf{\bibinfo{volume}{C79}},
  \bibinfo{pages}{044312} (\bibinfo{year}{2009}).

\bibitem[{\citenamefont{B\"{u}rvenich~{\it et al.}}(2002)}]{BMM.02x}
\bibinfo{author}{\bibfnamefont{T.}~\bibnamefont{B\"{u}rvenich~{\it et al.}}},
  \bibinfo{journal}{Phys. Rev.} \textbf{\bibinfo{volume}{C65}},
  \bibinfo{pages}{044308} (\bibinfo{year}{2002}).

\bibitem[{\citenamefont{Zdu{\'n}czuk~{\it et al.}}(2007)}]{ZSDK.07x}
\bibinfo{author}{\bibfnamefont{H.}~\bibnamefont{Zdu{\'n}czuk~{\it et al.}}},
  \bibinfo{journal}{Phys. Rev.} \textbf{\bibinfo{volume}{C76}},
  \bibinfo{pages}{044304} (\bibinfo{year}{2007}).

\end{thebibliography}

\end{document}